\documentclass[twocolumn,preprintnumbers,amsmath,amssymb]{revtex4-2}

\usepackage{graphicx}  
\usepackage{sidecap}
\begin{document}
\title{
 Coulomb electron drag mechanism of terahertz plasma instability  in n$^+$-i-n-n$^+$ graphene  FETs with ballistic injection
}
\author{V. Ryzhii$^{1,2,a}$,   M. Ryzhii$^3$,   V. Mitin$^4$, M. S. Shur$^5$, and T.~Otsuji$^1$}
\address{
$^1$Research Institute of Electrical Communication, Tohoku University, Sendai 980-8577, Japan\\
$^2$Institute of Ultra High Frequency Semiconductor Electronics of RAS,
 Moscow 117105, Russia\\
 $^3$Department of Computer Science and Engineering, University of Aizu, Aizu-Wakamatsu 965-8580, Japan\\
$^4$Department of Electrical Engineering,~University at Buffalo,~SUNY, Buffalo,~ New York 14260 USA\\
$^5$Department of Electrical,~Computer,~and~Systems~Engineering, Rensselaer Polytechnic Institute,~Troy,~New York 12180, USA\\
$^{a}$Author to whom correspondence should be addressed: v-ryzhii@riec.tohoku.ac.jp
}
 \begin{abstract} 
We predict the self-excitation of terahertz (THz) oscillations due to the plasma instability
in  the lateral  n$^+$-i-n-n$^+$ graphene   field-effect transistors (G-FET). 
The instability is associated with the Coulomb drag of the quasi-equilibrium electrons in the gated  channel by the injected ballistic electrons 
resulting in a positive feedback between the   amplified dragged electrons current and the injected current.  The plasma excitations   arise when the drag effect is sufficiently strong. The drag efficiency and the plasma frequency are determined by the quasi-equilibrium electrons  Fermi energy
(i.e., by their density). The  conditions
of the terahertz plasma oscillation self-excitation can be realized in the G-FETs with realistic structural parameters at room temperature enabling
the  potential   G-FET-based radiation sources  for the THz applications.
\end{abstract} 
\maketitle
%


\begin{figure}[t]
\centering
\includegraphics[width=7.0cm]{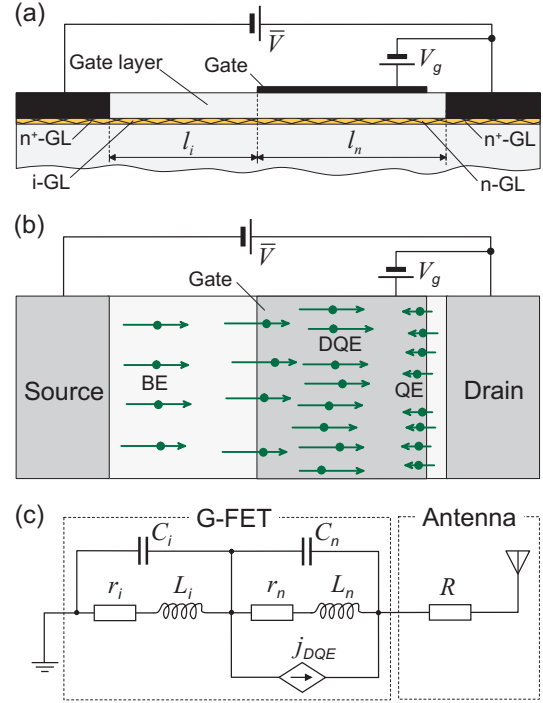}
\caption{(a) Cross-section  and (b) top view of a lateral  n$^+$-i-n-n$^+$  G-FET 
with ballistic i-region and gated n-region using the Coulomb electron drag
and (c) G-FET  equivalent circuit including an antenna with radiation resistance $R$.
}
\label{F1}
\end{figure}

The self-excitation of high-frequency plasma oscillations in the  field-effect transistor (FET) structures~\cite{1}
associated with the plasma instability was predicted a long time ago~\cite{2,3}.
 The detection and emission of the terahertz (THz) radiation in FETs, primarily attributed to the plasma oscillations excitation, 
was 
reported in many theoretical and experimental papers (see~\cite{4} and the references therein).
The plasmonic effects in graphene field-effect transistors (G-FETs)~\cite{5} can enable 
 marked advantages of such devices for the THz detection and emission~\cite{6,7,8,9,10,11,12,13,14}.

The linearity of the carrier energy dispersion law in graphene determines
the specifics of the electron-electron (electron-hole and
hole-hole) Coulomb interactions. In particular, it enables
a pronounced  Coulomb carrier drag~\cite{15,16,17,18,19,20,21} and can affect
the  plasma-electron-beam instability~\cite{22}.
As demonstrated recently~\cite{23}, in the lateral n$^+$-i-n-n$^+$
G-FETs with the structure shown in Figs.~1(a) and 1(b), the two-dimensional
ballistic electrons (BEs) injected from the source via the i-region
into the gated
n-region  can effectively drag the quasi-equilibrium electrons
(QEs) in this region. Due to  the BE-QE scattering, the current of the dragged QEs can exceed the current of the injected BEs (the current amplification, schematically shown in Fig.~1(c) as the current depended current source). The latter can cause
the reverse injection of the QEs from the drain and affect the G-FET current-voltage (IV) characteristics resulting in their strong nonlinearity~\cite{23}.

In this paper, we demonstrate that the QE drag by the BEs in the G-FETs can provide
the positive feedback between the BE current injected from the source and the QE reverse current. This can enable the instability of the DC current flow in the G-FET channel. Since the QE system in the G-FET can play the role of the THz resonant plasmonic cavity~\cite{3,5}, the instability in question can lead to the self-excitation of the THz  plasma oscillations feeding an antenna
and  resulting in the THz radiation.
 We derive the G-FET source-drain  small-signal impedance as a function of the device structural parameters and show that the real part of the  G-FET impedance can be negative
at the plasma  frequency, at which the imaginary part turns zero, that corresponds to the self-excitation of the plasma  oscillations~\cite{24}.

We assume that:\\
(i) The length, $l_i$, of the  undoped i-region (see Fig.~1(a)) is  sufficiently short allowing for the ballistic motion of the injected electrons: $l_i \ll \tau_{i}/v_W$, where 
$v_W \simeq 10^8$~cm/s,  $\tau_{i}$ is the characteristic time of the BE scattering on the acoustic phonons and impurities. In the graphene encapsulated in hBN, this condition can be realized if $l_i \lesssim 1~\mu$m at room temperature~\cite{25} and in even fairly long graphene layers at reduced temperatures~\cite{26};\\
(ii) The characteristic time, $\tau_{ee}$,  of the BE-QE collisions $\tau_{ee}$ is the shortest scattering time in the n-region: $\tau_{ee} < \tau_{op} \ll \tau_{n}$, where the latter times are associated with the scattering of the BEs on the optical phonons and  other momentum-non-conserving  collisions  such  as  those  involving impurities;\\
(iii) Then the energy of the BEs injected into the n-region exceeds the energy of optical phonons ($\hbar\omega_0 \simeq 200$~meV), the pertinent scattering mechanism can substantially dissipate the energy and momentum of the BEs.\\
 
When the length of the gated n-region  $l_n > v_W\tau_{ee}$, the majority of the BEs
injected into the n-region transfer their energy and, what is crucial, the momentum
to the QEs.  This causes their drag toward the drain forming the current exceeding the injected current since  despite the energy conservation 
at the electron-electron collisions, the net velocity of the electron system can increase.
At the electron densities $\Sigma_n \simeq 1\times(10^{12} - 10^{13})$~cm$^{-2}$ and  temperature $T \lesssim 300$~K for the  energy, $\varepsilon_{BE}$,  of the BEs injected into the n-region about the optical phonon energy $ \hbar\omega_0 \simeq 200$~meV one can set  $\tau_{ee}^{-1} \simeq (10 - 50)$~ps$^{-1}$ and   
 $\tau_{op}^{-1} = (1-2)$~ps$^{-1}$~\cite{27} with  
 the QE net scattering time $\tau_n$ on the acoustic phonons and impurities
about   $\tau_{n} \sim 1 - 2$~ps~\cite{28}. 
The minimization of  the BE scattering on impurities in the n-region is possible due to primarily  electricstatic doping of this region by the gate voltage.
Similar situation can occur in the case of the n-region  doping by donors  placed sufficiently far away from the channel when $\tau_{ee}$ is much shorter than the time of the BE collisions with the remote charged impurities.

 Considering the G-FET with the equivalent circuit shown in Fig.~1(c)
 and equalizing the BE current across the i-region and the net currents across
 the n-region, we arrive at the following equation:

\begin{equation}\label{eq1}
j_{BE} + j_{DP}^{(i)} = j_{DQE} + j_{QE} + j_{DP}^{(n)}.
 \end{equation}
Here  $j_{BE}$ and  $j_{QE}$ are the densities of the BE and QE
currents  across the i-region and the gated n-region, accounting for
their resistances and kinetic inductivities,  $j_{DP}^{(i)} =   c_id (\Phi_i)/dt$ and
$j_{DP}^{(n)} =   c_nd (V -\Phi_i)/dt$ are the displacement currents, 
$\Phi_i(t)$ and $V(t) - \Phi_i(t)$ are the potential drops across the i- and the n-regions, respectively. The capacitances are given by $C_i/H =c_i = (\kappa /2\pi^2)\ln\lambda$ (accounting for the specific of the structure~\cite{29,30,31,32}) and $C_n/H =c_n = l_n\kappa/2\pi\,d$, 
where $\kappa$ is the dielectric constant of the material in which graphene is embedded, $\ln \lambda \simeq \ln(4l_n/l_i)$ is on the order of unity, $H$ is the G-FET width, and  
$d$ is  the gate layer thickness.

\begin{figure}[t]
\centering
\includegraphics[width=8.0cm, bb=0 0 429 539]{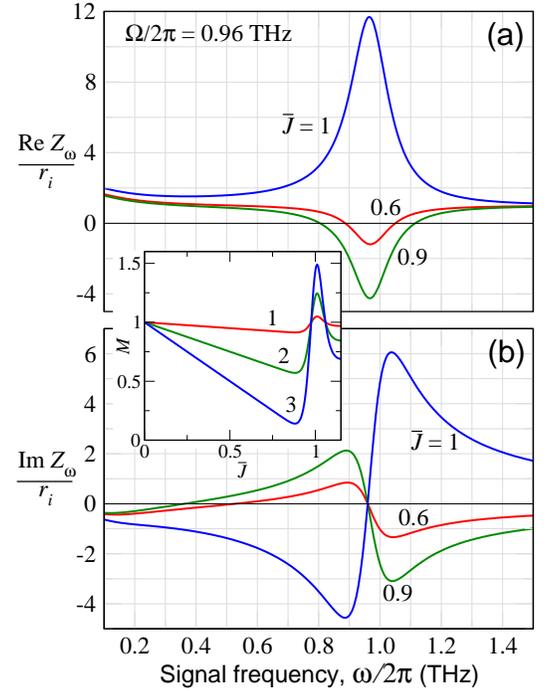}
\caption{The   real part Re~$Z_{\omega}/r_i$ (a) and  imaginary part Im~$Z_{\omega}/r_i$ (b) of the G-FET impedance
versus  signal frequency $f =\omega/2\pi$
for different values of normalized bias current ${\overline J}$: 
$\Omega/2\pi =0.96$~THz, $\tau_n = 1$~ps, $\mu_n = 50$~meV, $b = 1.47$, $\eta = 9.2$, and $ R/r_i = 1$. Inset shows parameter $M$ versus  bias current ${\overline J}$ for different values of $2b/(1+\eta)$:
1 - $2b/(1+\eta) = 0.1$, 2 - 0.5, and 3 - 1.0.
}
\label{F2}
\end{figure}

\begin{figure}[t]
\centering
\includegraphics[width=8.0cm,  bb=0 0 435 533]{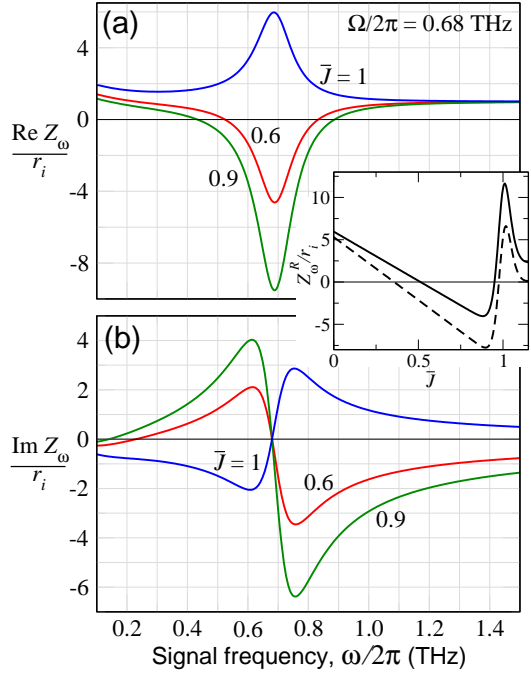}
\caption{The same as in Figs.~2(a) and 2(b), but for
$\Omega/2\pi =0.68$~THz, $\tau_n = 1$~ps, $\mu_n = 30$~meV, $b = 2.25$, $\eta = 5.52$, and $ R/r_i = 1$. Inset shows the resonant impedance $Z_{\omega}^R/r_i$ 
versus
  normalized bias current ${\overline J}$ for different values of the plasma frequency:  
$\Omega/2\pi =0.96$~THz - solid line and  $\Omega/2\pi = 0.68$~THz - dashed line
(other parameters are the same as for Fig.~2 and Fig.3, respectively).
}
\label{F3}
\end{figure}

 The 
 drag current density is given by ~\cite{23} 
 $j_{DQE} = bj_{BE}^2e^{-K}/j_0$, 
where   
$j_0 = v_W(\kappa\hbar\omega_0/2\pi\,l_ie)$ is the characteristic current density,
$b = (\hbar\omega_0/\mu_n)e^{-K_{n}}$ is the Coulomb drag factor
(which describes the drag current multiplication) with $K_n =l_n/v_W\tau_n$. The exponent
$K = \displaystyle K_{op}[(\varepsilon_{BE} + \mu_n)/\hbar\omega_0 -1]\cdot \Theta(\varepsilon_{BE}/\hbar\omega_0-1)$
is the probability of the optical phonon emission by a BE with
the energy $\varepsilon_{BE}$, which accounts for
the Pauli principle and for 
 the threshold character of such an emission process characterized by the step-like function $\Theta(z)$. The function $\Theta(Z) = [1+\exp(-2z\hbar\omega_0/k_BT]^{-1}$ describes  the temperature 
 smearing of the optical phonon emission threshold ($k_BT$ is the Boltzmann constant).

At the dc bias voltage $V = {\overline V}$ applied between the source and drain contacts and the dc voltage drop $\overline \Phi$ across the i-region,
the source-drain current density is equal to ${\overline j_{BE}} =   \sigma_i{\overline \Phi}_i/l_i$ and $\overline j_{QE} = \sigma_n ({\overline V}- {\overline \Phi})/l_n$. Here $\sigma_i = v_W\kappa/2\pi$ is the i-region  dc conductivity in the "virtual cathode" approximation~\cite{29} and $\sigma_n = e^2\Sigma_n \tau_n/m_n$ is the drift (Drude) conductivity of the n-region, where $\Sigma_n$, $\tau_n$, and $n_m$ are the QE  density, scattering time, and fictitious effective mass.
In this case, considering that $\varepsilon_{BE}/\hbar\omega_0  = e{\overline \Phi}_i/ \hbar\omega_0$  ($e$ is the electron charge) and introducing the normalized current density ${\overline J = {\overline j_{BE}}/j_0}$,
from Eq.~(1) we arrive at the following  equation relating 
${\overline J}$ and~${\overline V}$:

\begin{eqnarray}\label{eq2}
 {\overline J}-  
\frac{b}{(1+\eta)}{\overline J}^2e^{-K({\overline J})} = \frac{\eta}{(1+\eta)}\frac{{\overline V}}{V_0}.
\end{eqnarray}
Here $\eta = \sigma_nl_i/\sigma_il_n$ is the ratio of the i- and n-regions resistances $r_i = l_i/\sigma_iH$, $r_n= l_n/\sigma_nH$, and  $V_0 = \hbar\omega_0/e$. Equation~(2) describes the monotonic and the S-shaped IV characteristics at $2b/(1+\eta) < 1$ and $2b/(1+\eta) >1$, respectively~\cite{23}.

Considering the G-FET dynamic response, 
we assume 
that the voltages $V$ and $\Phi_i$ comprise the ac components: $V = {\overline V} + \delta V_{\omega} \exp(-i\omega t)$ and $\Phi_i = {\overline \Phi}_i + \delta \Phi_{\omega}$, where $\omega$ is the signal frequency, and
the normalized source-drain current also includes the pertinent ac contribution $\delta J_{\omega} = \delta \Phi_{\omega}/V_0$.

In this case, in the linearized version of Eq.~(1) we put $\delta j_{BE}/j_0 = \sigma_i\delta \Phi_{\omega}/l_ij_0(1+i\omega\tau_i) = \delta \Phi_{\omega}/V_0(1+i\omega\tau_i)$ and $\delta j_{QE}/j_0 = \sigma_n(\delta V_{\omega} - \delta\Phi_{{\omega}})/l_nj_0(1+ i\omega\tau_n) =\eta(\delta V_{\omega} - \delta\Phi_{{\omega}})/V_0(1+ i\omega\tau_n)$,
where $\tau_i$ and $\tau_n$  determine  the pertinent regions kinetic inductance. The scattering time $\tau_n$ coincide with the ratio of the n-region inductance and resistance.
Since the  transit time of the BEs across the i-region is short, the i-region
 kinetic ballistic inductance can be disregarded.
We also disregard the displacement current across the i-region due to $c_i \ll c_n$. This is justified in the range of frequencies under consideration.
 The voltage drop across the G-FET $\delta V_{\omega}$ can be expressed via the net ac voltage 
$\delta {\cal V}_{\omega}$, as 
 $\delta V_{\omega} = \delta {\cal V}_{\omega} - \delta {\overline J}_{\omega} j_0 RH$,
where  $R$  is the emitting antenna radiation resistance.

As a result, from Eq.~(1) for the ac component of the normalized current
$\delta {\overline J}_{\omega}$ we obtain the following equation:

\begin{eqnarray}\label{eq3}
\delta J_{\omega} = \frac{r_i}{Z_{\omega}}
\frac{\delta {\cal V}_{\omega}}{V_0}.
\end{eqnarray}
Here 
\begin{eqnarray}\label{eq4}
Z_{\omega} =r_i\,\Biggl[\frac{\displaystyle \frac{(M-M_0)}{M_0}(1+\omega^2\tau_n^2) }
{\displaystyle  1 +i\omega\tau_n\frac{(\Omega^2 - \omega^2 )\tau_n^2 - 1}
{\Omega^2\tau_n^2}} + 1 + \frac{R}{r_i}\Biggr]
\end{eqnarray}
 is the net impedance of the loop circuit under consideration. Deriving Eqs.~(3) and (4), we have introduced the quantities: 
$M_0 = \eta/(1+\eta)$,
$M =1 - [b/(1+\eta)]d[J^2e^{-K(J)}]/dJ|_{J ={\overline J} }$, 
which depend on the parameters $b$ and $\eta$, and
 the plasma frequency  

\begin{eqnarray}\label{eq5}
  \Omega =  \sqrt{\frac{8\pi\,e^2\Sigma_n\,d}{\kappa\,m_nl_n^2}}
= \frac{e}{\hbar\,l_n}\sqrt{\frac{8\mu_nd}{\kappa}}  
   \propto \frac{\sqrt{\mu_n}}{l_n}.
\end{eqnarray}
Here $\mu_n \simeq \hbar\,v_W\sqrt{\pi\Sigma_n}$ is the QE Fermi energy.
The  plasma frequency given by Eq.~(5) corresponds to its standard value for the plasma wavelength $\lambda = \sqrt{2}\pi\, l_n \simeq 4.4 l_n$.
Setting $\mu_n = 30 - 100$~meV, 
 $\kappa = 4$, $d = (5-10)\times 10^{-6}$~cm, and $l_n = 10^{-4}$~cm, we obtain
$\Omega/2\pi \simeq (0.53  -1.37)$~THz.

In the range of   low frequencies $\omega, \Omega \ll \tau_n^{-1}$, Eq.~(4)  yields
 $Z_{\omega} \simeq  R + r_i+r_n$ (in the absence of the Coulomb drug, $b = 0$ and $M = 1$) and $Z_{\omega} \simeq R + r_i$ (when the drug is pronounced, $M \ll 1$).

At the plasmonic resonance $\omega = \sqrt{\Omega^2 - \tau_n^{-2}}$,
the impedance imaginary part becomes zero, and Eq.~(4) yields

\begin{eqnarray}\label{eq6}
Z_{\omega}^R 
=  R  +r_i + r_i \frac{(M-M_0)}{M_0}\Omega^2\tau_n^2. 
\end{eqnarray}

Equation~(6) yields the condition $Z_{\omega}^R <0$ in the following forms:

\begin{eqnarray}\label{eq7}
\frac{M}{M_0} < 1 -\frac{(R + r_i)}{r_i\Omega^2\tau_n^2}, \qquad b> \frac{1 + \displaystyle\frac{(R + r_i)}{r_n}\frac{1}{\Omega^2\tau_n^2}}
{d [J^2e^{-K(J)}]/d J|_{J={\overline J}}}.
\end{eqnarray}
Considering that $\eta/\Omega^2\tau_n^2 = (l_il_n/4\pi\,dv_W\tau_n) =(l_i/4\pi\,d)K_n$, inequality~(7) can be presented as 

\begin{eqnarray}\label{eq8}
\frac{\hbar\omega_0}{\mu_n} > \frac{\biggl[1 +\displaystyle \biggl(\frac{ R}{r_i} + 1\biggr)\biggl(\frac{l_i}{4\pi\,d}\biggr)K_n\biggr]} 
{d [J^2e^{-K(J)}]/d J|_{J={\overline J}}} e^{\displaystyle K_n}.
\end{eqnarray}
The latter condition is valid at not too small $\mu_n$ ($\mu_n > k_BT$).  As follows from Eqs.~(7) and (8), the instability criteria primarily
requires a sufficiently large value $b = (\hbar\omega_0/\mu_n)e^{-K_n}$, i.e.,
not too large $K_n$. This implies a relatively strong Coulomb electron drag. 

Figures~2 and 3 show  the real part Re~$Z_{\omega}/r_i$ and  the imaginary part Im~$Z_{\omega}/r_i$ of normalized impedance versus signal frequency $\omega/2\pi$ calculated for different values of the normalized  bias current ${\overline J}$ using Eq.~(4) and the $M$ versus ${\overline J}$ dependence shown in the inset in Fig.2. 
The resonant impedance $Z_{\omega}^R/r_i$ as a function of the normalized bias current ${\overline J}$ calculated using Eq.~(6) is shown in the inset in Fig.~3. The structural parameters used for Figs.~2 and 3  correspond to realistic values: $l_i = 10^{-5}$~cm, $l_n = 10^{-4}$~cm, $d = 10^{-5}$~cm, $K_n = 1$, $K_{op} = 1$, and $\kappa = 4$.
For these parameters, assuming that the G-FET width  $H = (2 - 3)\times 10^{-3}$~cm, we obtain $r_i = 
[(\pi/4) - (\pi/6)]\times 10^{-10}$~s/cm  = 47 - 71~Ohm. This implies that at the realistic parameters of the G-FET structure, the resistance $r_i$
can match the standard antenna radiation resistance. 
When ${\overline J} \sim 1$, i.e., $\overline V \sim V_0 \simeq 200$~mV,
in a G-FET with the above parameters the dc current ${\overline J}j_0H \simeq 2.8$~mA.

One can see that at selected structural parameters and the bias current (bias voltage), Re~$Z_{\omega} <0$ in the THz range. Just in the range, where 
Re~$Z_{\omega} <0$, Im~$Z_{\omega}$ changes its sign turning zero at the plasmonic resonance. This corresponds to the self-excitation
of high-frequency  oscillations~\cite{24} - the plasma oscillations in our case, followed by  the  radiation emission from the antenna.

In conclusions,  we predicted the possibility of  the current driven plasma instability  in
the lateral G-FETs with the BE injection into the gated n-region region and the Coulomb drag of the QE by the BEs. The plasma instability and the pertinent self-excitation of the THz oscillation  are associated with the amplification
of the current due to the transfer of the BE momentum to the QEs.
The  plasma oscillations
self-excitation can lead to the THz radiation emission using the proper antenna. The G-FETs under consideration can be connected in series forming a periodic lateral structure (like in Ref.~\cite{11,33})  that  can enhance the THz emission.   

The work at RIEC and UoA was supported by the Japan Society for Promotion of Science (KAKENHI  Nos. 21H04546, 20K20349),
Japan; and the RIEC Nation-Wide Collaborative research
Project No. H31/A01, Japan.  The work at RPI was supported by the Office of Naval Research (N000141712976,
Project Monitor Dr. Paul Maki).

\section*{Data Availability}

The data that support the findings of this study are available
from the corresponding author upon reasonable request.

\end{document}